\documentclass[preprint,showpacs,preprintnumbers,amsmath,amssymb]{revtex4}

\usepackage{graphicx}
\usepackage{dcolumn}
\usepackage{bm}

\begin{document}

\title{New Spinor Field Realizations of the Non-Critical $W_{3}$ String}
\thanks{Supported by the National Natural Science Foundation of China under Grant No 10275030.}

\author{Li-Jie Zhang}
\affiliation{$^1$Department of Mathematics and Physics, Dalian Jiaotong University,
Dalian 116028}
\author{Yu-Xiao Liu}
\thanks{Corresponding author}\email{liuyx01@st.lzu.edu.cn}
\affiliation{$^2$Institute of Theoretical Physics, Lanzhou University, Lanzhou 730000}
\author{Ji-Rong Ren}
\affiliation{$^2$Institute of Theoretical Physics, Lanzhou University, Lanzhou 730000}

\begin{abstract}
We investigate the new spinor field realizations of the $W_{3}$ algebra,
making use of the fact that the $W_{3}$ algebra can be linearized by the addition of
a spin-1 current. We then use these new realizations to build the nilpotent Becchi-Rouet-Stora--Tyutin (BRST) charges
of the spinor non-critical $W_{3}$ string.
\end{abstract}

\pacs{11.25.Pm, 11.25.Sq, 11.10.-z.}

 \maketitle

As is well known, $W$ algebra has remarkable applications in $W$ gravity and $W$
string theories since its discovery in the 1980s.$^{[1,2]}$
Furthermore, it appears in the quantum Hall effect, black holes, in lattice models of
statistical mechanics at criticality, and in other physical models.$^{[3]}$

The  Becchi-Rouet-Stora--Tyutin (BRST)  formalism$^{[4]}$ has proven to be rather fruitful in the study of
the critical and non-critical $W$ string theories. The BRST charge of $W_{3}$
(i.e.~$W_{2,3}$) string was first constructed in Ref.\,[5], and its
detailed studies can be found in Refs.\,[5-7].
A natural generalization of the $W_{3}$ string, i.e. the $W_{2,s}$ strings, is a
higher-spin string with local spin-2 and spin-$s$ symmetries on the world-sheet. Many
works have been carried out on the scalar field realizations of $W_{2,s}$ strings.$^{[7-9]}$
The BRST charges for
such theories have been constructed for $s=4, 5, 6, 7$.$^{[8]}$
Later we
discovered the reason that the scalar BRST charge is difficult to be generalized to a
general $W_{N}$ string.$^{[10]}$ At the same time, we found the methods to
construct the spinor field realization of $W_{2,s}$ strings and $W_{N}$ strings.$^{[10]}$
Subsequently, we studied the exact spinor field realizations of
$W_{2,s}(s=3, 4, 5, 6)$ strings and $W_{N}(N=4, 5, 6)$ strings.$^{[10,11]}$
Recently, we have constructed the nilpotent BRST charges of spinor
non-critical $W_{2,s}(s=3,4)$ strings by taking into account the property of spinor field.$^{[12]}$
These results will be important for constructing super $W$
strings, and they will provide the essential ingredients.

However, all of these theories about the $W_{2,s}$ strings mentioned above are based on
the non-linear $W_{2,s}$ algebras. Because of the intrinsic nonlinearity of the $W_{2,s}$
algebras, their study is a more difficult task compared to linear algebras. Fortunately,
it has been shown that certain $W$ algebras can be linearized by the inclusion of a
spin-1 current. This provides a way of obtaining new realizations of the $W_{2,s}$
algebras. Such new realizations were constructed for the purpose of building the
corresponding scalar $W_{2,s}$ strings.$^{[13]}$

Since there has been no work focused on the research of spinor field realizations of the
non-critical $W_{3}$ string based on the linear $W_{1,2,3}$ algebra, in this Letter we construct
new nilpotent BRST charges of spinor non-critical $W_{3}$ string for the first time
by using the linear bases of the $W_{1,2,3}$ algebra. To construct a
non-critical BRST charge, one must first solve the forms of matter currents $T$ and $W$
determined by the OPEs of $TT$, $TW$ and $WW$. Here $T$ and $W$ here are constructed on the
linear bases $T_0, J_0, W_0$ of the $W_{1,2,3}$ algebra, and these linear bases are
constructed with the spinor fields. Then direct substitution of these results into BRST
charge leads to the grading spinor field realizations.
All these results will be important for embedding the
Virasoro string into the $W_{3}$ string.

We begin by reviewing the linearization of the $W_{3}$ algebra by the inclusion of a
spin-1 current.$^{[14]}$ We take the linearized $W_{1,2,3}$ algebra in the form
\begin{eqnarray}
T_{0}(z)T_{0}(\omega) &\sim & \frac{C/2}{(z - w)^4}+\frac{2T_{0}(\omega)}{(z-\omega)^2}+
\frac{\partial T_{0}(\omega)}{z-\omega}, \nonumber \\
 T_{0}(z)W_{0}(\omega) &\sim &
\frac{3 W_{0}(\omega)}{(z-\omega)^2}+ \frac{\partial
W_{0}(\omega)}{z-\omega}, \nonumber \\
T_{0}(z)J_{0}(\omega) &\sim &  \frac{C_{1}}{(z -
w)^3}+\frac{J_{0}(\omega)}{(z-\omega)^2}+ \frac{\partial
J_{0}(\omega)}{z-\omega}, \label{W123} \\
J_{0}(z)J_{0}(\omega) &\sim & \frac{-1}{(z-\omega)^2},\;\; \nonumber \\
J_{0}(z)W_{0}(\omega) &\sim&  \frac{hW_{0}(\omega)}{z-\omega},\;\;\; 
W_{0}(z)W_{0}(\omega) \sim  0.\nonumber
\end{eqnarray}
The coefficients $C$, $C_{1}$ and  $h$ are given by
\begin{eqnarray}\label{CC1h}
C = 50+24t^{2}+\frac{24}{t^{2}},\; 
C_{1} = -\sqrt{6}(t+\frac{1}{t}),\; 
h  =\sqrt{\frac{3}{2}} ~t.
\end{eqnarray}

To obtain the new realizations for the linearized $W_{1, 2, 3}$ algebra, we use the
multi-spinor fields $\psi^{\mu}$, which have spin 1/2 and satisfy the OPE
\begin{equation}
\psi^{\mu}(z)\psi^{\nu}(\omega) \sim - \frac{1}{z-\omega} ~\delta^{\mu\nu},\nonumber
\end{equation}
for the first time to construct the linear bases of them. The general forms of these
linear bases can be taken as follows:
\begin{eqnarray}
 T_0   = -\frac{1}{2} \partial \psi^{\mu} \psi^{\mu}, ~
 J_0   = \alpha_{\mu \nu} \psi^{\mu} \psi^{\nu} ~(\mu < \nu),~
 W_0   = 0,\nonumber
\end{eqnarray}
where $\alpha_{\mu \nu}$ are pending coefficients. By making use of the OPE
$J_{0}(z)J_{0}(\omega)$ in Eq.\,(1), we can obtain the equation that the coefficients
$\alpha _{\mu \nu}$ satisfy, i.e. $\sum _{\mu < \nu} \alpha _{\mu \nu}^2 =1$. From the
OPE relation of $T_{0}$ and $J_{0}$, it is easy to obtain $C_1=0$. Substituting the value of
$C_1$ into Eq.\,(2), we can obtain the value of $t$. Then the total central charge
$C$ for $T_0$ can be obtained from Eq.\,(2). Thus we can determine the explicit
form of $T_{0}$ under the restricted condition of its central charge. Finally, using the
OPE $T_{0}(z)J_{0}(\omega)$ in Eq.\,(1) again, we reach the exact form of $J_{0}$. The
complete results are $t=\pm i,~C=2,~C_{1}=0$ and
\begin{eqnarray}
T_0 = -\frac{1}{2} \sum_{\mu=1}^{4} \partial \psi^{\mu} \psi^{\mu},~ 
J_0 = \sum_{\mu < \nu = 1}^{4} \alpha_{\mu \nu} \psi^{\mu} \psi^{\nu},~ 
W_0 = 0,\label{W0spin3}
\end{eqnarray}
where the coefficients $\alpha_{\mu \nu}$ satisfy
\begin{eqnarray}
&\sum_{\mu < \nu =1}^{4} (\alpha_{\mu \nu})^2=1. \label{RestrictAlpha}
\end{eqnarray}

As we know, for the energy-momentum tensor of a scalar $\phi$, it has central charge $C=1$ in two-dimensional conformal field theory, while $C=1/2$ for a single fermion field. This means that one real scalar field is equivalent to two real or one complex fermion fields. In particular, they correspond to the forms $\psi=:e^{i\phi}:, \psi=\psi_1 + i \psi_2.$ Similarly, the power exponent can also be used as a means of bosonisation in our constructions. Then, our 4-fermion construction is equivalent to a two-scalar system.

Now let us consider the spinor realizations of the non-linear $W_{3}$ algebra with linear
bases of the $W_{1,2,3}$ algebra. We begin by reviewing the structures of the $W_{3}$
algebra in conformal language. The OPE $W(z)W(\omega)$ for $W_{3}$ algebra is given by$^{[1]}$

\begin{eqnarray}\label{OPEofWWspin3}
W(z)W(\omega) &\sim & \frac{C/3}{(z - w)^6} + \frac{2T}{(z-\omega)^4}
                      + \frac{\partial T}{(z-\omega)^3} \nonumber \\
                &+& \frac{1}{(z-\omega)^2}\left(2 \Theta \Lambda %
                            +\frac{3}{10}\partial ^2 T \right)  \\
                &+& \frac{1}{(z-\omega)}\left(\Theta \partial \Lambda %
                            +\frac{1}{15}\partial ^3 T\right), \nonumber
\end{eqnarray}
where
\begin{equation}
\Theta= \frac{16}{22+5C},\;\;\; \Lambda = T^2 - \frac{3}{10}
\partial ^2 T.\nonumber
\end{equation}

The bases of the $W_{3}$ algebra can be constructed by the linear bases of the
$W_{1,2,3}$ algebra:
\begin{eqnarray}
T = T_{0}, ~ 
W = W_{0}+W_{R}(J_{0},T_{0}),\nonumber
\end{eqnarray}
where the currents $T_{0}$, $J_{0}$ and $W_{0}$ generate the $W_{1,2,3}$ algebra and have
been constructed with multi-spinor fields. First we can write the most general
possible structure of $W$. Then the relations of the above OPEs of  $T$ and $W$ determine the
coefficients of the terms in $W$. Finally, substituting these coefficients together with $T_{0}$,
$J_{0}$ and $W_{0}$ into the expressions of $T$ and $W$, we can obtain the spinor
realizations of the $W_{3}$ algebra. The explicit results turn out to be very simple:
\begin{eqnarray}
T = T_0 , ~ 
W = W_{0} \pm \frac{1}{6}(- 3i \partial ^2 J_{0} + 4i J_{0}^{3}+ 6 i T_{0}J_{0}).
\label{TW_1}
\end{eqnarray}
Substituting the expressions of $T_{0}$, $J_{0}$ and $W_{0}$ in Eq. (3) into
Eq. (6), we obtain the explicit constructions of $T$ and $W$ for the $W_{3}$
algebra as follows:
\begin{eqnarray}
T &=& -\frac{1}{2} \sum_{\mu=1}^{4} \partial \psi^{\mu}
\psi^{\mu}, \label{Tspin3} \\
W &=& \pm \frac{1}{2i} \left(\sum_{\mu < \nu,
\lambda,\rho=1}^{4}|\epsilon_{\mu\nu\lambda\rho}| \alpha_{\mu\nu}
\psi^{\mu} \psi^{\nu} \partial\psi^{\lambda} \psi^{\lambda} \right.  \nonumber\\
&&\left.+ \sum_{\mu < \nu=1}^{4} \alpha_{\mu\nu} (\partial^2 \psi^{\mu} \psi^{\nu} +
\psi^{\mu}
\partial^2\psi^{\nu} + 2\partial\psi^{\mu} \partial\psi^{\nu})\right),\label{Wspin3}
\end{eqnarray}
where $\alpha_{\mu \nu}$ satisfies Eq.\,(4).

The non-critical $W_{3}$ string is the theory of $W_{3}$ gravity coupled to a matter
system on which the $W_{3}$ algebra is realized. Subsequently, we give realizations of the spinor field
.

The BRST charges of the non-critical $W_{3}$ string take the form:
\begin{eqnarray}
    Q_{B}&=&Q_{0}+Q_{1},\label{QB}\\
    Q_{0}\;&=&\oint dz\; c(T_{\psi}+T_{M}+T_{bc}+yT_{\beta\gamma}+T^{eff}),\label{Q0}\\
    Q_{1}\;&=&\oint dz\; \gamma F(\psi,\beta,\gamma,T_{M},W_{M}),\label{Q1}
\end{eqnarray}
where $y$ is a pending constant. The matter currents $T_{M}$ and $W_{M}$, which have spin
2 and 3 respectively, generate the $W_{3}$ algebra and have been constructed for the
cases $s=3$ ($T$ and $W$ in Eqs.\,(7) and (8) are replaced by $T_{M}$ and
$W_{M}$, respectively). The energy-momentum tensors in Eq. (10) are given by$^{[12,15]}$
\begin{eqnarray}
    T_{\psi}&=&-\frac{1}{2}\partial\psi\psi,~
    T_{M}=T,~
    T_{bc}=2b\partial c+\partial bc,\nonumber\\
    T_{\beta\gamma}&=& 3\beta\partial\gamma+2\partial\beta\gamma,~
    T^{eff}=-\frac{1}{2}\eta_{\mu\nu}\partial Y^{\mu} Y^{\nu}.\nonumber
\end{eqnarray}
The operator $F(\psi,\beta,\gamma,T_{M},W_{M})$ has the spin of 3 and the ghost number of zero. The
BRST charge generalizes the one for scalar non-critical $W_{2,s}$ strings, and it is also
graded with $Q_{0}^{2}=Q_{1}^{2}=\{Q_{0},Q_{1}\}=0$. Since the first condition is
satisfied for any $s$ automatically, the remaining two conditions determine $y$ and the
coefficients of the terms in $F(\psi,\beta,\gamma,T_{M},W_{M})$.

Now, using the grading BRST method and the procedure mentioned above, we discuss the
exact solutions of spinor field realizations of the non-critical $W_{3}$ string.

In this case, $Q_{B}$ takes the form of Eq.\,(9).  The mater currents $T_{M}$ and
$W_{M}$ are given by Eqs.\,(7) and (8), respectively. The most
extensive combinations of $F$ in Eq.\,(11) with correct spin and ghost number can be
constructed as follows:
\begin{eqnarray}\label{Fspin3}
F &=&f_{1}\beta^3\gamma^3+f_{2}\partial\beta\beta\gamma^2
  + f_{3}\partial \beta \partial \gamma
  + f_{4} \beta\gamma\partial\psi\psi  \nonumber \\
  &&
  + f_{5} \beta \partial ^2\gamma
  + f_{6}\partial ^2\psi\psi  + f_{7}\beta\gamma T_{M}  \nonumber \\
  &&
  + f_{8}\partial\psi\psi T_{M}
  + f_{9}\partial T_{M} + f_{10}W_{M}.
\end{eqnarray}
Substituting Eq.\,(12) into Eq.\,(11) and imposing the nilpotency
conditions $Q_{1}^{2} = \{Q_{0},Q_{1}\}=0$, we can determine $y$ and
$f_{i}(i=1,2,\cdots,10)$. They correspond to three sets of general solutions, i.e.

\noindent (i) $y=0$,
$ f_{i}= 0~ (i=4,6,7,8,9,10),\nonumber $
and $f_{j}\;(j=1,2,3,5)$ are arbitrary constants but do not vanish at the same
time.

\noindent (ii) $y=1$ and
\begin{eqnarray}
&&f_{1} = \frac{1}{150}(-7 f_{3} + 3 f_{5}),~ 
f_{2} = \frac{1}{15}(7 f_{3} - 3 f_{5}),\nonumber \\
&&f_{4} = \frac{1}{5}(22 f_{3} - 78 f_{5} + 10 f_{7}),\nonumber \\
&&f_{6} = -11 f_{3} + 39 f_{5} - 5 f_{7},~ 
f_{8} = 0,~ f_{9} = -\frac{5}{2}f_{7},\nonumber
\end{eqnarray}
\noindent where $f_{j}\;(j=3,5,7,10)$ are arbitrary
constants but do not vanish at the same time.

\noindent (iii) $y$ is an arbitrary constant and
\begin{eqnarray}
f_{i} &=& 0\;\;\; (i=4,6,7,8,9,10), \nonumber \\
f_{1} &=& -\frac{8}{55}f_{5}, ~ f_{2} = \frac{16}{11} f_{5}, ~ f_{3} = \frac{39}{11}
f_{5}, \nonumber
\end{eqnarray}
where $f_{5}$ is an arbitrary non-zero constant.

Substituting the construction (12) of $F$ into Eq.\,(11) and using $T_{M}$ and
$W_{M}$ given by Eqs.\,(7) and (8), we obtain the explicit spinor field
realizations of the non-critical $W_{3}$ string.

In summary, we have constructed the new spinor field realizations of the non-critical $W_{3}$ string, making use of the fact that the $W_{3}$ algebra can be linearized by the addition of a spin-1 current. First, we use the multi-spinor fields $\psi^{\mu}$ to construct the linear bases of the linearized $W_{1, 2, 3}$ algebra. Subsequently, the non-linear bases of $W_{3}$ algebra is constructed with these linear bases $T_0, J_0$ and $W_0$. Finally, we use these new realizations to build the graded BRST charges of the spinor non-critical $W_{3}$ string. The constructions are based on demanding the nilpotency of the BRST charges. The solutions are very standard, i.e., there are three solutions for the $W_{3}$ string. It is worth noting that our four-fermion result has central charge $C=2$, while for the critical $W_3$ string the central charge should be $C=100$. Since the $(T_0,J_0)$ system is linear, we can add more matter to obtain $C=100$. For simplicity, considering the independence of central charge under non-critical case, we only give the discussion based on $C=2$ in this study. We expect that there should exist such realizations for the case of higher spin $s$. Having obtained the exact BRST charges of $W_{3}$ string, we can investigate the implications for the corresponding string theories.

It is a pleasure to thank Professor DUAN Yi-Shi and Dr. WEI Hao for useful discussions. One
of the authors (LIU Yu-Xiao) thanks Professor LU Jian-Xin and
Professor CAI Rong-Gen for their suggestive discussions and hospitality.

\end{document}